 %
\input harvmac
%
%
 %
\catcode`@=11
\def\rlx{\relax\leavevmode}                   
 %
 %
 %
\font\eightrm=cmr8 \font\eighti=cmmi8 \font\eightsy=cmsy8
  
\skewchar\eighti='177 \skewchar\eightsy='60
%

 %
\font\tenmib=cmmib10
\font\sevenmib=cmmib10 at 7pt 
\font\fivemib=cmmib10 at 5pt  
\font\tenbsy=cmbsy10
\font\sevenbsy=cmbsy10 at 7pt 
\font\fivebsy=cmbsy10 at 5pt  
\def\BMfont{\textfont0\tenbf \scriptfont0\sevenbf
                              \scriptscriptfont0\fivebf
            \textfont1\tenmib \scriptfont1\sevenmib
                               \scriptscriptfont1\fivemib
            \textfont2\tenbsy \scriptfont2\sevenbsy
                               \scriptscriptfont2\fivebsy}
\def\BM#1{\rlx\ifmmode\mathchoice
                      {\hbox{$\BMfont#1$}}
                      {\hbox{$\BMfont#1$}}
                      {\hbox{$\scriptstyle\BMfont#1$}}
                      {\hbox{$\scriptscriptstyle\BMfont#1$}}
                 \else{$\BMfont#1$}\fi}
 %
 %
 %
 %
\def\inbar{\vrule height1.5ex width.4pt depth0pt}
\def\sinbar{\vrule height1ex width.35pt depth0pt}
\def\ssinbar{\vrule height.7ex width.3pt depth0pt}
\font\cmss=cmss10
\font\cmsss=cmss10 at 7pt
\def\ZZ{\rlx\leavevmode
             \ifmmode\mathchoice
                    {\hbox{\cmss Z\kern-.4em Z}}
                    {\hbox{\cmss Z\kern-.4em Z}}
                    {\lower.9pt\hbox{\cmsss Z\kern-.36em Z}}
                    {\lower1.2pt\hbox{\cmsss Z\kern-.36em Z}}
               \else{\cmss Z\kern-.4em Z}\fi}
\def\Ik{\rlx{\rm I\kern-.18em k}}  
\def\IC{\rlx\leavevmode
             \ifmmode\mathchoice
                    {\hbox{\kern.33em\inbar\kern-.3em{\rm C}}}
                    {\hbox{\kern.33em\inbar\kern-.3em{\rm C}}}
                    {\hbox{\kern.28em\sinbar\kern-.25em{\sevenrm C}}}
                    {\hbox{\kern.25em\ssinbar\kern-.22em{\fiverm C}}}
             \else{\hbox{\kern.3em\inbar\kern-.3em{\rm C}}}\fi}
\def\IP{\rlx{\rm I\kern-.18em P}}
\def\IR{\rlx{\rm I\kern-.18em R}}
\def\Ione{\rlx{\rm 1\kern-2.7pt l}}
 %
 %

 %

\def\intem#1{\par\leavevmode%
              \llap{\hbox to\parindent{\hss{#1}\hfill~}}\ignorespaces}
 %


 %
\newskip\humongous \humongous=0pt plus 1000pt minus 1000pt   
\def\caja{\mathsurround=0pt}
\newif\ifdtup
 %
\def\cmath#1{\,\vcenter{\openup2\jot \caja
     \ialign{\strut \hfil$\displaystyle{##}$\hfil\crcr#1\crcr}}\,}
 %
\def\eqalign#1{\,\vcenter{\openup2\jot \caja
     \ialign{\strut \hfil$\displaystyle{##}$&$
      \displaystyle{{}##}$\hfil\crcr#1\crcr}}\,}
 %
\def\twoeqsalign#1{\,\vcenter{\openup2\jot \caja
     \ialign{\strut \hfil$\displaystyle{##}$&$
      \displaystyle{{}##}$\hfil&\hfill$\displaystyle{##}$&$
       \displaystyle{{}##}$\hfil\crcr#1\crcr}}\,}
 %
\def\panorama{\global\dtuptrue \openup2\jot \caja
     \everycr{\noalign{\ifdtup \global\dtupfalse
      \vskip-\lineskiplimit \vskip\normallineskiplimit
      \else \penalty\interdisplaylinepenalty \fi}}}
 %

 %
\def\eqalignno#1{\panorama \tabskip=\humongous
     \halign to\displaywidth{\hfil$\displaystyle{##}$
      \tabskip=0pt&$\displaystyle{{}##}$\hfil
       \tabskip=\humongous&\llap{$##$}\tabskip=0pt\crcr#1\crcr}}
 %

 %
\def\twoeqsalignno#1{\panorama \tabskip=\humongous
     \halign to\displaywidth{\hfil$\displaystyle{##}$
      \tabskip=0pt&$\displaystyle{{}##}$\hfil
       \tabskip=0pt&\hfil$\displaystyle{##}$
        \tabskip=0pt&$\displaystyle{{}##}$\hfil
         \tabskip=\humongous&\llap{$##$}\tabskip=0pt\crcr#1\crcr}}
 %

 %
 %
 %
 %
   \let\SS=\S       
\def\,{\hskip1.5pt}           
 %
\let\a=\alpha
\let\b=\beta

\let\d=\delta       \let\vd=\partial             
\let\e=\epsilon     
                       \let\F=\Phi
\let\g=\gamma                                    
\let\h=\eta

\let\j=\psi                                      

                                   \let\L=\Lambda

                         \let\P=\Pi
\let\q=\theta                   \let\Q=\Theta
         
\let\s=\sigma       \let\vs=\varsigma            \let\S=\Sigma

\let\x=\xi                                       \let\X=\Xi
                                  \let\Y=\Upsilon

 %
 %
\def\Box{{\sqcap\mkern-12mu\sqcup}}
\def\lapp{\lower.4ex\hbox{\rlap{$\sim$}} \raise.4ex\hbox{$<$}}
\def\gapp{\lower.4ex\hbox{\rlap{$\sim$}} \raise.4ex\hbox{$>$}}
\def\con{\ifmmode\raise.1ex\hbox{\bf*}
          \else\raise.1ex\hbox{\bf*}\fi}

\let\To=\Rightarrow

\def\Imm{\mathop{\Im m}}

\def\dual{\relax\leavevmode\lower.9ex\hbox{\titlerms*}}
\def\define{\buildrel\rm def\over =}
\let\id=\equiv
\let\8=\otimes
 %
 %
 %
 %

\let\2=\underline

 %
\def\dt#1{{\buildrel{\smash{\lower1pt\hbox{.}}}\over{#1}}}
\def\pd#1#2{{\partial#1\over\partial#2}}

\def\6(#1){\relax\leavevmode\hbox{\eightrm(}#1\hbox{\eightrm)}}
\def\0#1{\relax\ifmmode\mathaccent"7017{#1}     
                \else\accent23#1\relax\fi}      
\def\7#1#2{{\mathop{\null#2}\limits^{#1}}}      
\def\5#1#2{{\mathop{\null#2}\limits_{#1}}}      
 %

 %

 %

 %

 %
\newbox\t@b@x
\def\rightarrowfill{$\m@th \mathord- \mkern-6mu
     \cleaders\hbox{$\mkern-2mu \mathord- \mkern-2mu$}\hfill
      \mkern-6mu \mathord\rightarrow$}
\def\tooo#1{\setbox\t@b@x=\hbox{$\scriptstyle#1$}%
             \mathrel{\mathop{\hbox to\wd\t@b@x{\rightarrowfill}}%
              \limits^{#1}}\,}
\def\leftarrowfill{$\m@th \mathord\leftarrow \mkern-6mu
     \cleaders\hbox{$\mkern-2mu \mathord- \mkern-2mu$}\hfill
      \mkern-6mu \mathord-$}
\def\froo#1{\setbox\t@b@x=\hbox{$\scriptstyle#1$}%
             \mathrel{\mathop{\hbox to\wd\t@b@x{\leftarrowfill}}%
              \limits^{#1}}\,}
 %
\def\frac#1#2{{#1\over#2}}
\def\frc#1#2{\relax\ifmmode{\textstyle{#1\over#2}} 
                    \else$#1\over#2$\fi}           
\def\inv#1{\frc{1}{#1}}                            
 %
\def\Claim#1#2#3{\bigskip\begingroup%
                  \xdef #1{\secsym\the\meqno}%
                   \writedef{#1\leftbracket#1}%
                    \global\advance\meqno by1\wrlabeL#1%
                     \noindent{\bf#2}\,#1{}\,:~\sl#3\vskip1mm\endgroup}

\def\QED{\rlx\hfill$\Box$\kern-7pt\raise3pt\hbox{$\surd$}\bigskip}
 %
 %
\def\1{\raise1pt\hbox{,}}     
\def\Tr{\mathop{\rm Tr}}
\def\:{\buildrel!\over=}
\def\ex#1{\hbox{$\>{\rm e}^{#1}\>$}}
\def\CP#1{\rlx\ifmmode\IP^{#1}\else\IP$^{#1}$\fi}
\def\cP#1{\rlx\ifmmode\IC{\rm P}^{#1}\else$\IC{\rm P}^{#1}$\fi}

\def\sll#1{\rlx\rlap{\,\raise1pt\hbox{/}}{#1}}
\def\Sll#1{\rlx\rlap{\,\kern.6pt\raise1pt\hbox{/}}{#1}\kern-.6pt}
\let\sss=\scriptscriptstyle
\let\SSS=\scriptstyle
\let\ttt=\textstyle

 %
\def\ie{\hbox{\it i.e.}}        
 %

\def\CY{Calabi-\kern-.2em Yau}

\def\3{\ifmmode\ldots\else$\ldots$\fi}
\def\\{\hfill\break}
\def\Z{\hfil\break\rlx\hbox{}\quad}
\def\3{\ifmmode\ldots\else$\ldots$\fi}
\def\ping{\nobreak\par\centerline{---$\circ$---}\goodbreak\bigskip}
 %
 %

 %

 %
\def\JP#1{{\it J.\,Phys.\,}{\bf#1\,}}

\def\NP#1{{\it Nucl.\,Phys.\,}{\bf#1\,}}
\def\PL#1{{\it Phys.\,Lett.\,}{\bf#1\,}}

\def\CQG#1{{\it Class.\,Quant.\,Grav.\,}{\bf#1\,}}

\baselineskip=13.0861pt plus2pt minus1pt
\parskip=\medskipamount
\let\ft=\foot
\noblackbox
 %
\def\Afour{\ifx\answ\bigans
            \hsize=16.5truecm\vsize=24.7truecm
             \else
              \hsize=24.7truecm\vsize=16.5truecm
               \fi}
 %
 %
\def\SaveTimber{\abovedisplayskip=1.5ex plus.3ex minus.5ex
                \belowdisplayskip=1.5ex plus.3ex minus.5ex
                \abovedisplayshortskip=.2ex plus.2ex minus.4ex
                \belowdisplayshortskip=1.5ex plus.2ex minus.4ex
                \baselineskip=12pt plus1pt minus.5pt
 \parskip=\smallskipamount
 \def\ft##1{\unskip\,\begingroup\footskip9pt plus1pt minus1pt\setbox%
             \strutbox=\hbox{\vrule height6pt depth4.5pt width0pt}%
              \global\advance\ftno by1
               \footnote{$^{\the\ftno)}$}{\ninepoint##1}%
                \endgroup}}
\catcode`@=12
%
%
%

\SaveTimber  
 %
 %
\def\rd{{\rm d}}

\def\Zp{\big|}
\def\?{\,\hbox{--}}

\def\@{{\ttt\cdot}}
\def\T{{}^{\sss\rm T}}
\def\pp{{\mathchar'75\mkern-9mu|\mkern3mu}} 
\def\mm{{=}}                                
\def\Pp{{\id\mkern-9.5mu|\mkern3mu}}        
\def\Mm{{\id}}                              

\def\B#1{\mathaccent"7616{#1}}
\def\]#1{\mkern#10mu}
\def\[#1{\mkern-#10mu}
\def\N{\nabla}
\def\bN{\overline\nabla}
 %

\def\ad{{\dot\a}}

\def\bA{{\bf A}}
\def\bAb{\relax\leavevmode\hbox{{\bf A}\kern-.7em
               \vrule height1.9ex depth-1.8ex width5pt}\kern2pt}

\def\bd{{\dot\b}}

\def\bB{{\bf B}}
\def\bBb{\relax\leavevmode\hbox{{\bf B}\kern-.7em
               \vrule height1.9ex depth-1.8ex width5pt}\kern2pt}

\def\Cb{\relax\leavevmode\hbox{$C$\kern-.53em
               \vrule height1.9ex depth-1.8ex width5pt}\kern.5pt}
\def\bC{{\bf C}}
\def\bCb{\relax\leavevmode\hbox{{\bf C}\kern-.65em
               \vrule height1.9ex depth-1.8ex width5pt}\mkern2mu}
\def\CC{\BM{\cal C}}
\def\bCC{\relax\leavevmode\hbox{\BM{\cal C}\kern-.5em
               \vrule height1.9ex depth-1.8ex width5pt}\kern.8pt}
\def\db{{\mathaccent"7616\partial}}

\def\Db{\relax\leavevmode\hbox{$D$\kern-.6em
               \vrule height1.9ex depth-1.8ex width5pt}\kern.5pt}

\def\cDb{\relax\leavevmode\hbox{$\cal D$\kern-.6em
               \vrule height1.9ex depth-1.8ex width5pt}\kern.5pt}

\def\rDb{\relax\leavevmode\hbox{D\kern-.7em
               \vrule height1.9ex depth-1.8ex width5pt}\kern1.5pt}
\def\CE{\BM{\cal E}}
\def\bCE{\relax\leavevmode\hbox{\BM{\cal E}\kern-.5em
               \vrule height1.9ex depth-1.8ex width4pt}\kern.8pt}

\def\Fb{\relax\leavevmode\hbox{$F$\kern-.55em
               \vrule height1.9ex depth-1.8ex width4.5pt}\kern1pt}
\def\FB{\relax\leavevmode\hbox{$\Phi$\kern-.6em
               \vrule height1.9ex depth-1.8ex width4.5pt}\kern1.5pt}
\def\bF{{\bf F}}

\def\gd{{\dot\g}}

\def\Gb{\relax\leavevmode\hbox{$\Gamma$\kern-.55em
               \vrule height1.9ex depth-1.8ex width4.5pt}\kern1pt}

\def\GB{{\bf\Gamma}}
\def\bGB{\overline{\bf\Gamma}}
\def\iGb{\relax\leavevmode\hbox{$\mit\Gamma$\kern-.55em
               \vrule height1.9ex depth-1.8ex width4.5pt}\kern1pt}
\def\cG{{\cal G}}
\def\CG{\BM{\cal G}}

\def\CH{\BM{\cal H}}
\def\bCH{\relax\leavevmode\hbox{\BM{\cal H}\kern-.7em
               \vrule height1.9ex depth-1.8ex width6pt}\kern.8pt}

\def\JB{\relax\leavevmode\hbox{$\Psi$\kern-.7em
               \vrule height1.9ex depth-1.8ex width6pt}\kern1.5pt}
\def\CK{\BM{\cal K}}
\def\bCK{\relax\leavevmode\hbox{\BM{\cal K}\kern-.7em
               \vrule height1.9ex depth-1.8ex width6pt}\kern.8pt}


\def\LB{\relax\leavevmode\hbox{$\Lambda$\kern-.6em
               \vrule height1.9ex depth-1.8ex width5pt}\kern1.5pt}

\def\Mb{\relax\leavevmode\hbox{$M$\kern-.8em
               \vrule height1.9ex depth-1.8ex width7pt}\kern.8pt}

\def\cO{{\cal O}}

\def\pB{\relax\leavevmode\hbox{$\BMfont p$\kern-.45em
               \vrule height1.35ex depth-1.25ex width4pt}\kern1pt}
\def\Pb{\relax\leavevmode\hbox{$P$\kern-.55em
               \vrule height1.9ex depth-1.8ex width4.5pt}\kern.5pt}
\def\PB{\relax\leavevmode\hbox{$\P$\kern-.6em
               \vrule height1.9ex depth-1.8ex width4.5pt}\kern.5pt}
\def\qb{\mathaccent"7616\theta}
\def\qB{\relax\leavevmode\hbox{$\BMfont q$\kern-.45em
               \vrule height1.35ex depth-1.25ex width4pt}\kern1pt}

\def\Qb{\relax\leavevmode\hbox{$Q$\kern-.6em
               \vrule height1.9ex depth-1.8ex width5.5pt}\kern.5pt}
\def\QB{\relax\leavevmode\hbox{$\Q$\kern-.65em
               \vrule height1.9ex depth-1.8ex width5.0pt}\kern.5pt}

\def\rQb{\relax\leavevmode\hbox{Q\kern-.65em
               \vrule height1.9ex depth-1.8ex width5pt}\kern1.5pt}
\def\bQb{\relax\leavevmode\hbox{{\bf Q}\kern-.65em
               \vrule height1.9ex depth-1.8ex width5pt}\kern1.5pt}

\def\vsb{\mathaccent"7616\varsigma\mkern2mu}

\def\Tb{\relax\leavevmode\hbox{$T$\kern-.55em
               \vrule height1.9ex depth-1.8ex width4.5pt}\kern.5pt}

\def\bVb{\relax\leavevmode\hbox{{\bf V}\kern-.65em
               \vrule height1.9ex depth-1.8ex width5pt}\kern1.5pt}
\def\CV{\BM{\cal V}}
\def\Wb{\relax\leavevmode\hbox{$W$\kern-.9em
               \vrule height1.9ex depth-1.8ex width7pt}\kern1.5pt}
\def\bW{{\bf W}}
\def\bWb{\relax\leavevmode\hbox{{\bf W}\kern-.95em
               \vrule height1.9ex depth-1.8ex width7pt}\kern1.5pt}
\def\CW{\BM{\cal W}}

\def\Xb{\relax\leavevmode\hbox{$X$\kern-.67em
               \vrule height1.9ex depth-1.8ex width6pt}\kern.5pt}
\def\bXB{\relax\leavevmode\hbox{\BM{X}\kern-.77em
               \vrule height1.9ex depth-1.8ex width6pt}\kern1.5pt}
\def\XB{\relax\leavevmode\hbox{$\X$\kern-.65em
               \vrule height1.95ex depth-1.85ex width6pt}\kern.5pt}
\def\YB{\relax\leavevmode\hbox{$\Upsilon$\kern-.6em
               \vrule height1.9ex depth-1.8ex width4.5pt}\kern1.5pt}

\def\ssl#1{\rlx\rlap{\,\raise1pt\hbox{$\backslash$}}{#1}}
\def\Ssl{\rlap{\kern1.2pt\raise1pt\hbox{\rm/}}{\hbox{$S$}}}

 %

 %

 %
 %
 %
\Title{\rightline{hep-th/0002112}}
      {\vbox{\centerline{Yang-Mills and Supersymmetry Covariance}
              \vskip3mm
             \centerline{Must Coexist}}}
\centerline{\titlerms Tristan H\"ubsch\footnote{$^{\spadesuit}$}{On leave
            from the ``Rudjer Bo\v skovi\'c'' Institute, Zagreb, Croatia.}}
                                                             \vskip0mm
 \centerline{\it Department of Physics and Astronomy}        \vskip-.5mm
 \centerline{\it Howard University, Washington, DC~20059}    \vskip-.5mm
 \centerline{\tt thubsch\,@\,howard.edu}
\vfill

\centerline{ABSTRACT}\vskip2mm
\vbox{\narrower\narrower\baselineskip=12pt\noindent
 Supersymmetry and Yang-Mills type gauge invariance are two of the
essential properties of most, and possibly the most important models in
fundamental physics. Supersymmetry is nearly trivial to prove in the
({\it traditionally\/} gauge-noncovariant) superfield formalism,
whereas the gauge-covariant formalism makes gauge invariance manifest. In
3+1-dimensions, the transformation from one into the other is elementary and
essentially unique. By contrast, this transformation turns out to be {\it
impossible\/} in the most general 1+1-dimensional case. In fact, only the
(manifestly) gauge- and supersymmetry-covariant formalism guarantees both
universal gauge-invariance and supersymmetry.}
 \vskip+5mm
 \rightline{{\ninepoint\it There are cracks in every building;}}
 \rightline{{\ninepoint\it that's how the light comes in.}}
 \rightline{{\ninepoint\it|~L.~Cohen}}

\Date{02/00. \hfill}  
\footline{\hss\tenrm--\,\folio\,--\hss}
 %
 %
 %
\lref\rBK{I.L.~Buchbinder and S.M.~Kuzenko: {\it Ideas and Methods of
        Supersymmetry and\Z Supergravity : Or a Walk Through Superspace},
        (IOP Publishing, Bristol, 1998).}

\lref\rTwJim{S.J.~Gates, Jr.: \PL{B352}(1995)43--49.}

\lref\rGGRS{S.J.~Gates, Jr., M.T.~Grisaru, M.~Ro\v cek and
       W.~Siegel: {\it Superspace}\Z (Benjamin/Cummings Pub.\ Co.,
       Reading, Massachusetts, 1983).}

\lref\rGGW{S.J.~Gates, Jr., M.T.~Grisaru and M.E.~Wehlau:
       \NP{B460}(1996)579--614.}

\lref\rGHR{S.J.~Gates, Jr., C.M.~Hull and M.~Ro\v cek: \NP{B248}(1984)157.}

\lref\rHPS{C.M.~Hull, G.~Papadopoulos and B.~Spencer:
       \NP{B363}(1991)593-621.}

\lref\rBeast{T.~H\"ubsch: {\it \CY\ Manifolds---A Bestiary for
      Physicists}\Z (World Scientific, Singapore, 1992).}

\lref\rHSS{T.~H\"ubsch: \NP{B555}(1999)567-628.}

\lref\rGauSS{R.Q.~Almukahhal and T.~H\"ubsch: Gauging Yang-Mills Symmetries
      In 1+1-Dimensional Spacetime. hep-th/9910007.}

\lref\rChiLin{T.~H\"ubsch: \CQG{16}(1999)L51-L54.}

\lref\rIvan{E.A.~Ivanov: \JP{A16}(1983)2571.}

\lref\rWB{J.~Wess and J.~Bagger: {\it Supersymmetry and Supergravity}\Z
      (Princeton University Press, Princeton NJ, 1983).}

\lref\rPW{P.~West: {\it Introduction to Supersymmetry and Supergravity}\Z
      (World Scientific, Singapore, 1990).}

\lref\rWAB{E.~Witten: 
      in {\it Essays on Mirror Manifolds}, p.120, Ed.~S.-T.~Yau
      (International Press, Hong Kong, 1992).}

\lref\rPhases{E.~Witten: \NP{B403}(1993)159--222.}

 %
 %
\newsec{Introduction}\noindent
\seclab\sIntro
Both global (rigid) and local (gauged) symmetry play an important r\^ole in
physics. Local `internal' (Yang-Mills gauge) symmetry provides for
understanding of all fundamental interactions except gravity, while the
supersymmetric extension of spacetime (`external') symmetries provides for
the so essential stability under quantum fluctuations, and the only known
link between `internal' and `external' symmetries.
 It is then vexing to find most of the existing literature fail to treat
these two essential symmetries in a {\it simultaneously\/} covariant
fashion~\refs{\rGGRS,\rBK,\rGauSS}.

The purpose of the present article is then to explore this, first
briefly in the familiar $N{=}1$ supersymmetric 3+1-dimensional spacetime,
and then more extensively in the $(2,2)$-supersymmetric 1+1-dimensional
one. In both cases, the `misalignment' between Yang-Mills and supersymmetry
covariance provides a geometrical interpre\-tation---indeed, a rigorous
definition---of the prepotential superfields~\refs{\rGGRS,\rBK,\rGauSS}. In
the familiar $N{=}1$ supersymmetric 3+1-dimensional spacetime, the
(de)covariantizing transformation between the two formalisms is well
known and essentially unique~\refs{\rGGRS,\rBK}. By contrast, and perhaps
not surprisingly~\refs{\rGHR,\rHPS,\rTwJim}, in the (2,2)-supersymmetric
1+1-dimensional case, we find {\it two\/} distinct `(de)covariantizing'
transformations which toggle between the gauge-covariant and the more
traditional {\it simple\/} (i.e., gauge-noncovariant)
framework\ft{Throughout, `simple' (`simply') will be used to mean
`gauge-{\it non\/}covariant(ly)', as opposed to `gauge-covariantly'.
Also, supersymmetry will be treated in a manifestly covariant
fashion.}. One of them simplifies matters about gauge-covariantly
(anti)chiral superfields but complicates dealing with non-minimal
gauge-covariantly twisted-(anti)chiral superfields, while the other
achieves precisely the opposite. Both decovariantizing transformations,
however, complicate matters regarding gauge-covariantly unidexterous
(lefton and righton) and all but one of the non-minimally (NM-)haploid
superfields~\refs{\rGGRS,\rBK,\rGauSS,\rHSS}. Therefore, it is impossible
to turn all gauge-covariantly haploid superfields {\it simultaneously\/}
into simply haploid superfields.

This article is organized as follows: Section~2 recalls the notion,
notation and use of gauge-covariant superderivatives, and defines the
various gauge superfields. The familiar case of $N{=}1$ supersymmetric
3+1-dimensional spacetime is briefly reviewed in Section~3, which also
presents a swift and elementary derivation of the geometrical origin of the
gauge prepotential superfield, complementing the literature~\rIvan; see
also \SS\,3.6.3 of Ref.~\rBK. Section~4 then explores this in
(2,2)-supersymmetric 1+1-dimensional spacetime. In particular, we find it
{\it impossible\/} to ensure universal gauge-invariance for every model
constructed in the simple (gauge-noncovariant) formalism~\rHSS\ by the
`standard' judicious insertion of $\ex{2\CV}$-like terms~\refs{\rBK,\rWB}.
By contrast, it is straightforward to gauge-covariantize (and so render
gauge-invariant) all of them by gauge-covariantizing all
(super)derivatives\rGGRS.

\newsec{Gauge-Covariant Superderivatives}\noindent
\seclab\sSSC
Local (gauge) symmetry affects parallel transport and so changes all
(super)derivatives into gauge-covariant (super)derivatives. This in turn
modifies the supersymmetry algebra, and is the starting point for our
analysis following
\SS\,4.2.b of~\rGGRS\ and \SS\,3.6.4 of~\rBK.

\subsec{Universal (super)derivatives}\noindent
\subseclab\ssDefs
Adjusting the notation so as to conform with Refs.~\refs{\rWB,\rPhases},
we follow Ref.~\rGGRS\ and start by defining the {\it covariant\/}
(super)derivatives
\eqn\eCDs{{\cmath{
 \N_\a=D_\a-i\GB_\a~, \qquad \bN_\ad=\Db_\ad-i\bGB_\ad~, \cr
 \N_m=\vd_m-i\GB_m~, }}}
where $\a$ ($\ad$) label the components of the (co-)spinors, and $m$
those of the vectors of the appropriate Lorentz symmetry group.
 The $\N$'s representing the geometrical concept of parallel transport,
their definition~\eCDs\ must be {\it universal\/}. That is, all
(super)fields transforming under the same gauge symmetry must couple to the
same gauge (super)fields appearing in the definitions~\eCDs, and in the
same way. Equivalently, the coupling of gauge (super)fields to
matter (super)fields only depends on the charges of the latter. The use of
the term `{\it universal\/} gauge-covariance' is herein meant to emphasize
this universalty in coupling of gauge and matter (super)fields.

 The gauge (connexion form coefficient) superfields $\GB,\bGB$ are gauge
algebra valued. For nonsimple gauge groups, $\cG=\prod_I\cG_I$, the $\GB$'s
are linear combinations of gauge superfields each of which is valued in the
gauge algebra of only one factor, $\cG_I$. Gauge coupling parameters are
suppressed and are easily reinserted by replacing $\GB\to g\GB$.

 Recall that the superderivatives~\refs{\rGGRS,\rBK,\rWB}
\eqn\eDs{ D_\a\define\vd_\a - i\qb^\ad\s^m_{\a\ad}\vd_m \qquad
          \Db_\ad\define-\vd_\ad + i\q^\a\s^m_{\a\ad}\vd_m }
already include the `$\inv2$-form connexions',
$-i\qb^\ad\s^m_{\a\ad}\vd_m$ and $+i\q^\a\s^m_{\a\ad}\vd_m$, valued in the
spacetime translation group generators, $\vd_m$. The (fermionic) `gauge
superfields' $-i\qb^\ad\s^m_{\a\ad}$ and $+i\q^\a\s^m_{\a\ad}$ however
being constant in spacetime, supersymmetry is rigid (global).

\subsec{Gauge transformations}\noindent
\subseclab\ssGTrs
Gauge transformations act linearly on `matter' superfields
\eqna\eGTr
 $$
    \BM{X}' ~=~ \CG\BM{X}~, \eqno\eGTr{a}
 $$
implemented by the {\it unitary\/} operator superfield $\CG$:
 $$
    \bXB' ~=~ \bXB\CG^{\dag} ~=~ \bXB\CG^{-1}~. \eqno\eGTr{b}
 $$
It is with respect to this gauge transformation that the
(super)derivatives~\eCDs{} are covariant, so that (suppressing spacetime
indices):
\eqn\eCov{ \N' ~=~ \CG\,\N\,\CG^{-1}~. }

Being a gauge group element, $\CG$ can be written as
\eqn\eGOp{ \CG~=~\ex{i\CE}~,\qquad \CE\>\define\CE^iT_i~,\qquad
           [T_j,T_k]~=~if_{jk}{}^lT_l~, }
where $T_i$ are the (hermitian) generators of the gauge group with structure
constants $f_{jk}{}^l$. $\CE$ is gauge algebra valued and must be Hermitian
for $\CG$ to be unitary.

 The transposition in~\eGTr{b} makes the action of gauge-covariant
derivatives on $\bXB$ awkward: the derivative part of $\N$ should act from
the left as usual, but the gauge superfield (connexion form) part should act
from the right. We therefore calculate (implicitly) using a double
transposition~\rBK:
\eqn\eTTr{ (\cO\bXB)~\define~(\cO\T\bXB\T)\T~
                      =~(\B\cO\BM{X})^\dagger~, }
where $\cO$ denotes {\it any\/} gauge-covariant operator, $\cO\T$ its
transpose, and $\B\cO\id\cO^\dagger$ its {\it Hermitian\/} conjugate. In
practice, and in cases when above `matrix' notation would be ambiguous or
confusing, we resort to the explicit gauge group index notation. With the
matter fields forming a representation of the gauge group the
elements of which are indexed by $\a,\b,\3$, Eqs.~\eGTr{}--\eCov\
and~\eTTr\ become:
\eqn\eGTR{ \BM{X}'{}^\a ~=~ \CG_\b^\a\BM{X}^\b~, \qquad
           \bXB_\a' ~=~ \bXB_\b\CG^{\dag}{}_\a^\b
                       ~=~ \bXB_\b\CG^{-1}{}_\a^\b~, }
\eqn\eCOV{ \N'_\a{}^\b ~=~ \CG_\a^\g\,\N_\g{}^\d\,[\CG^{-1}]_\d^\b~, }
and
\eqn\eTTrInd{ (\cO\bXB)_\a~\define~(\cO^\b_\a\bXB_\b)
              ~=~(\B\cO{}^\a_\b\BM{X}^\b)^\dagger~, }
respectively. This disentangles ordering issues and the `matrix' action of
the gauge fields on the matter fields: re-ordering now solely depends on
the spin/statistics of the involved superfields and operators. In this
article, explicit gauge group indices will be suppressed, hoping that the
Reader will always be able to discern the implied meaning of the more
compact `matrix' notation.

\subsec{Gauge superfields}\noindent
\subseclab\ssGSFs
Under the gauge transformation~\eCov, the gauge superfields, $\GB$,
transform {\it inhomogeneously\/}:
\eqn\eInh{ \GB' = \CG\,\GB\,\CG^{-1} -i\CG^{-1}(D\CG)~. }
Suitable choices of $\CG$ then produce a cancellation on the right hand
side for some of the component fields of $\GB$. All such (partially fixed)
choices of $\CG$ are generally called Wess-Zumino gauges, and depend on the
desired number and structure of component field cancellations on the right
hand side of $\GB$.

 In contrast, field strength superfields and torsion superfields, $\bF$ and
$\bf T$, are defined by (anti)commutation of the covariant
derivatives~\eCDs{}, according the master formula\ft{Here ``$[~{,}~\}$''
denotes the (anti)commutator, as appropriate for the (anti)commuted
quantities.} (omitting spacetime indices and conjugation bars)
\eqn\eFTs{ \big[\, \N \,,\, \N \,\big\}~=~{\bf T}{\cdot}\N -i\bF~, }
which determines $\bF,\bf T$ in terms of the gauge superfields $\GB$ upon
using Eqs.~\eCDs{}. By virtue of their definition~\eFTs, field strength
superfields, $\bF$, and torsion superfields, $\bf T$, are {\it covariant\/}:
they transform {\it homogeneously\/} with respect to the gauge
transformation $\CG$. Therefore, the vanishing of some of the component
fields of a torsion or field strength superfield, $\bf T$ or $\bF$, is a
gauge-invariant statement.

This then provides the only way to unambiguously constrain the superfield
content in a supersymmetric model with gauge symmetry: constraints to this
end involve only torsion and field strength superfields, and so are
gauge-invariant statements. Finally, the use of (gauge-covariantly
constrained) superfields guarantees the maintenance of both supersymmetry
and gauge-covariance.

\newsec{Yang-Mills Symmetry in $(3,1\,|\,1)$-Superspacetime}\noindent
\seclab\sFour
The gauge-covariant superderivatives~\eCDs\ close a modified supersymmetry
algebra which has been discussed in the literature~\refs{\rGGRS,\rBK}:
\eqna\eTRS
 $$\eqalignno{
 \{\N_\a,\bN_\ad\}&\define2i\s^m_{\a\ad}\N_m~,                &\eTRS{a}\cr
 \{\N_\a,\N_\b\}&=~0~=~\{\bN_\ad,\bN_\bd\}~,                  &\eTRS{b}\cr
 }$$
stating, respectively, that ${\bf T}_{\a\ad}=2i\s^m_{\a\ad}$ and
$\bF_{\a\ad}=0$, and that ${\bf T}_{\a\ad}=0=\bF_{\a\b}$.
The vanishing of the field strength, 
$\bF_{\a\ad}$, on the right hand side of~\eTRS{a} ensures no duplication of
gauge fields per gauge transformation by enforcing
\eqn\eVecG{ \GB_m = \frc{i}4\s_m^{\a\ad}
 \big(\{D_\a,\bGB_\ad\}+\{\GB_\a,\Db_\ad\}-i\{\GB_\a,\bGB_\ad\}\big)~. }
 The vanishing of~\eTRS{b} ensures that gauge-covariantly
chiral superfields (see below) can couple to the gauge superfields~\rGGRS.
Jacobi identities then further ensure that:
\eqn\eDefW{ [\bN_\ad,\N_{\b\bd}] = \e_{\ad\bd}\bW_\b~,
            \quad\hbox{where}\quad
            \N_{\a\ad}\define\s^m_{\a\ad}\N_m~, }
\eqn\eChiW{ \bN_\ad\bW_\a~=~0~, \quad\hbox{and}\quad
            \N^\a\bW_\a+ \bN^\ad\bWb_\ad~=~0~, \quad\hbox{and} }
\eqn\eTheF{ \bF_{\a\ad,\b\bd}~\define~i[\N_{\a\ad},\N_{\b\bd}]~=~
 (\e_{\a\b}\bN_{(\ad}\bWb_{\bd)}+\e_{\ad\bd}\N_{(\a}\bW_{\b)})~, }
where $\bF_{\a\ad,\b\bd}=\s^m_{\a\ad}\s^n_{\b\bd}\bF_{mn}$, so that
the gauge-covariantly chiral spin-$\inv2$ superfields $\bW_\a,\bWb_\ad$
encode all the gauge field strengths.

Gauge-covariantly chiral and antichiral superfields are then defined to
satisfy
\eqn\eGChi{ \bN_\ad\F~=~0~,\qquad \N_\a\FB~=~0~, }
whereas gauge-covariantly complex linear superfields and their conjugates
satisfy
\eqn\eGLin{ \bN^\ad\bN_\ad\Q~=~0~,\qquad \N^\a\N_\a\QB~=~0~, }
respectively. From Eq.~\eGChi, it easily follows that
\eqn\eGCInt{ 0~=~\{\bN_\ad,\bN_\bd\}\F~
             =~{\bf T}_{\ad\bd}{\cdot}\N\F+\bF_{\ad\bd}\F~. }
This implies that $\bF_{\ad\bd}=0$ or $\F$ would have to be chargeless.
Also, ${\bf T}_{\ad\bd}{}^\a=0$ and ${\bf T}_{\ad\bd}{}^m=0$ so $\F$ would
not be forced to be a constant. Finally, the component fields of
${\bf T}_{\ad\bd}{}^\gd\id{\bf T}_{(\ad\bd)}{}^\gd$ include spin-$\frc32$
fields (the traceless part) and the appearance of $\bN_\gd$ on the right
hand side of the~\eGCInt, as a special case of~\eFTs, would induce a further
modification of Eqs.~\eCDs; ${\bf T}_{(\ad\bd)}{}^\gd$ is therefore
also set to zero, justifying~\eTRS{}.

\subsec{Misalignment of gauge- and super-symmetry}\noindent
\subseclab\ssMisAl
In the gauge-covariant framework, the fermionic integration is expressed as
an appropriate gauge-covariant superderivative, followed by the $\q,\qb\to0$
projection. For example~\rTwJim,
\eqn\eCDI{ \hbox{D-term:}\qquad \int\rd^4\q~K
 ~=~\inv8\Tr\big[\{\N^\a\N_\a,\bN^\ad\bN_\ad\}\,K\big]\Zp~, }
and in similar vein:
\eqn\eCFI{ \hbox{F-term:}\qquad
            \int\rd^2\q~W~=~\inv2\Tr\big[\N^\a\N_\a\,W\big]\Zp~, }
where `$|$' denotes setting $\q,\qb=0$. The
gauge-covariance~\eGTr{}--\eCov\ of the constraints~\eGChi\ and~\eGLin\
makes the expansion of fermionic integral like~\eCDI\ and~\eCFI\
straightforward, if perhaps tedious to evaluate.

Proving supersymmetry of the fermionic integrals~\eCDI\ and~\eCFI\ is
practically trivial. The (gauge-covariantized) supersymmetry
transformation operator may be rewritten as
\eqn\eSuSy{ \d_{\e,\B\e}\>\define\e{\cdot}{\cal Q}+\B\e{\cdot}\B{\cal Q}
            ~=~\e{\cdot}\!\!\int\rd\q+\B\e{\cdot}\!\!\int\rd\qb
             +4\x^m\vd_m~+\h^i(\e,\B\e;\GB,\bGB)\,T_i~, }
where $\x^m=\Imm(\e{\cdot}\s^m{\cdot}\qb)$. The quantities
$\h^i$ are linear combinations of $\e,\B\e$ and functions of $\GB,\bGB$, the
exact form and value of which depend on the precise choice of ${\cal
Q},\B{\cal Q}$\ft{The gauge-covariantization, ${\cal Q},\B{\cal Q}$, of the
supercharges, $Q,\Qb$, is constrained by the requiring that
 $\{\N_\a,{\cal Q}_\b\}=0=\{\bN_\ad,{\cal Q}_\b\}$, and
 $\{{\cal Q}_\a,\B{\cal Q}_\ad\}=-2i\s^m_{\a\ad}\N_m$. Therefore, if
$\GB,\bGB$ gauge (non)\-abelian symmetries, the $\h^i$ will be (non)linear
functions of $\GB,\bGB$.}. Notice that all the Lagrangian densities,
such as~\eCDI\ and~\eCFI, are gauge-invariant, and so must be annihilated
by the gauge group generators,
$T_i$. Being $\q,\qb$-independent, they are also annihilated by the
Berezin superintegration operators.

 Therefore, the supersymmetry transformations of the Lagrangian
densities~\eCDI\ and~\eCFI\ must be proportional, respectively, to the total
derivatives
\eqn\eTotD{ \vd_m\big[\{\N^\a \N_\a,\bN^\ad\bN_\ad\}\,\x^mK\big]\Zp
 \quad\hbox{and}\quad \vd_m\big[\N^\a \N_\a\,\x^mW\big]\Zp~. }
Finally: the spacetime integrals of these expressions (the supersymmetry
variations of the corresponding terms in the action) vanish owing to the
usual assumptions of sufficiently rapid vanishing at `infinity' of all
involved fields in open spacetimes, and by virtue of Stokes' theorems in
closed spacetimes.

\subsec{The decovariantizing transformation}\noindent
\subseclab\ssDeCov
Since the gauge-covariant superderivatives~\eCDs\ transform
covariantly~\eCov, there is of course no gauge in which the $\N,\bN$'s would
become `ordinary' superderivatives, $D,\Db$~\eDs. More precisely, no {\it
unitary\/} operator can possibly transform the $\N,\bN$'s into the
$D,\Db$'s.

 There do exist, however, {\it nonunitary\/} operators $\CH$ such that
\eqn\eDeCov{ \N_\a = \bCH^{-1}D_\a\bCH~, \quad\hbox{and}\quad
           \bN_\ad = (\N_\a)^{\dagger} = \CH\Db_\ad\CH^{-1}~. }
The expansion~\eCDs\ provides the identification
\eqn\eGinH{ \GB_\a~=~+i\bCH^{-1}(D_\a\bCH)~,\qquad
           \bGB_\ad~=~-i(\Db_\ad\CH)\CH^{-1}~. }
Through Eq.~\eVecG, this also determines $\GB_m$. Therefore, all gauge
superfields are expressible in terms of $\CH$ and its Hermitian conjugate.
Being non-unitary, $\CH$ is an element of the {\it complexified\/} gauge
group, $\cG^c$. Written as an exponential, $\CH$ has a complex exponent, the
antihermitian part of which can be removed by a compensating gauge
transformation, so that, in this `Hermitizing' gauge,
\eqn\eHGTr{ \CH~\to~\CH^g=\bCH^g=\ex{\CV}, \qquad \CV^{\dag}=\CV~. }
Therefore, $\CV$ generates the coset $\cG^c/\cG$, locally at the particular
(unitary, `Hermitizing') gauge transformation which turned $\CH$ Hermitian.

According to Eq.~\eFTs, the gauge-covariant (dynamical and physically
relevant!) field strengths are, through the substitution~\eDeCov,
determined as {\it second\/} (super)derivatives of $\CH,\bCH$ and their
inverses. It then follows that the lowest and the next-to-lowest
components of $\CH,\bCH$|and so also $\CV$|are unphysical, and are
eliminated in suitable so-called Wess-Zumino
gauge(s)~\refs{\rGGRS,\rWB,\rBK}; see \SS\,\ssGSFs.

The non-unitary transformation of gauge-covariant derivatives~\eDeCov\
induces a corresponding transformation of gauge-covariantly constrained
superfields. The chirality condition~\eGChi\ becomes
\eqn\eXXX{ 0~=~\bN_\ad\F=\CH\Db_\ad\CH^{-1}\F~, }
which begs for the conjugate pair of definitions
\eqn\eNChi{ \0\F~\define~\CH^{-1}\F~, \quad\hbox{and}\quad
            \0\FB~\define~\FB\bCH^{-1}~. }
Easily, $\0\F$ and $\0\FB$ are chiral and antichiral with respect
to the `ordinary' superderivatives~\eDs:
\eqn\eXXX{{\eqalign{
 0=\bN_\ad\F=\CH \Db_\ad\0\F &~~\To~~\Db_\ad\0\F=0~, \cr
 0=\N_\a\T\FB\T=(\bCH^{-1}D_\a\bCH)\T\FB\T=\bCH\T D_\a\0\FB\T
 &~~\To~~D_\a\0\FB=0~, \cr }}}
Similarly, for gauge-covariantly complex linear superfields we define
\eqn\eNLin{ \0\Q~\define~\CH^{-1}\Q~, }
where $\0\Q$ is complex linear with respect to the `ordinary'
superderivatives~\eDs:
\eqn\eXXX{ 0=\bN^2\Q=\CH \Db^2\0\Q ~~\To~~\Db^2\0\Q=0~. }
The analogous is true of its Hermitian conjugate, $\0\QB$.

{\bf Remark}: the condition $\bN_\ad=(\N_\a)^{\dagger}$, implicit in~\eCov\
and made explicit in~\eDeCov, may be relaxed, so that $\bN_\ad$ (defining
$\F$) and $\N_\a$ (defining $\FB$; see below) are decovariantized
independently and differently. This implies a {\it separate\/}
decovariantizing transformation rule for the gauge-covariantly chiral and
antichiral superfields, $\F,\FB$, and so leads to a group of transformations
larger than $\cG$, or even $\cG^c$. The present discussion is
straightforward albeit unnecessary to extend in this way, and we do not do
so herein.

\subsec{Lagrangian density}\noindent
\subseclab\ssLagD
The decovariantizing transformation, of course, also changes Lagrangian
density terms. For example,
\eqn\eNDI{{\eqalign{
 \Tr\big[\N^2\bN{}^2\T\,\F\FB\big]\Zp
 &=\Tr\big[\bCH^{-1}D^2\bCH (\CH\Db^2\CH^{-1})\T
            \CH\0\F\0\FB\bCH\big]\Zp~,\cr
 &\to\Tr\big[\CH^{-1}D^2\CH \CH^{-1}\Db^2\CH
            \CH\0\F\0\FB\CH\big]\Zp~,\cr
 &\]2=D^2\Db^2\>\Tr[\0\FB\ex{2\CV}\0\F]\Zp~.\cr}}}
Here the second line follows upon taking the Hermitizing gauge~\eHGTr, and
we used the cyclicity of the trace operation to obtain the last line. The
same is true of the other half of the expression~\eCDI. In fact, the same
result is obtained with a general D-term of the form $\int\rd^4\q\,K$, as
long as the real function $K=K(\F,\FB,\Q,\QB)$ transforms as $K=\CH
\0K \bCH$, and where of course, $\0K\define K(\0\F,\0\FB,\0\Q,\0\QB)$.
 So, the (seemingly) `flat' kinetic D-term in the gauge-covariant framework
becomes the standard gauge-invariant kinetic D-term in the decovariantized
framework and upon the Hermitizing gauge transformation~\eHGTr. This clearly
identifies $\CV\cong\log(\CH)$, the local generator of the $\cG^c/\cG$
coset, as the gauge prepotential superfield: Eqs.~\eGinH\ become
\eqn\eGinV{ \GB_\a~=~+i(D_\a\CV)~,\qquad
           \bGB_\ad~=~-i(\Db_\ad\CV)~. }
See Ref.~\rIvan\ and \SS\,3.6.3 of Ref.~\rBK\ for a derivation
of these facts from the `opposite' vantage point. 

As well known, the F-term $\Tr\big[\N^2\,W\big]\Zp$ is
gauge-invariant only if the superpotential $W$ is. Then $[T_i,W]=0$ and
also $[\CG,W]=0=[\CH,W]$, so that
\eqn\eNFI{ \Tr\big[\N^2\,W\big]\Zp~=~\Tr\big[\CH^{-1}D^2\CH\,W\big]\Zp~
          =~D^2\,\Tr[\0W]\Zp~, }
using the cyclicity of the trace operation, and the gauge invariance of
$W=\0W$. Notice that the trace operator may well be omitted in the
end expressions of~\eNDI\ and~\eNFI.

 In all cases, supersymmetry of the expressions in Eqs.~\eNDI\
and~\eNFI\ is straightforward, by virtue of the simple proof given in
\SS\,\ssMisAl, involving Eqs.~\eSuSy\ and~\eTotD. Thus, the supersymmetry
of the `gauge-decovariantized' Lagrangian density terms~\eNDI\ and~\eNFI\
is still {\it manifest\/}, whereas gauge-covariance is marred by the use of
the simple superderivatives~\eDs, both in Berezin integration~\eNDI\
and~\eNFI\ and the definition of constrained superfields~\eNChi\ and~\eNLin.

At present, in the $N{=}1$ supersymmetric 3+1-dimensional spacetime, this
`misalignment' may seem a matter of (in)convenience, although it does
identify the gauge prepotential superfield, $\CV$, as a local generator
of the $\cG^c/\cG$ coset.

\newsec{Yang-Mills Symmetry in $(1,1\,|\,2,2)$-Superspacetime}\noindent
\seclab\sTwo
Turn now to the $(2,2)$-supersymmetric 1+1-dimensional spacetime. Let
$\s^\mm{\define}\inv2(\s^0{-}\s^3)$ and $\s^\pp{\define}\inv2(\s^0{+}\s^3)$
be the light-cone characteristic (bosonic) coordinates, and
$\vs^\mp,\vsb^\mp$ the fermionic coordinates. The spinor indices
$\a,\ad=-,+$ in fact denote spin: $\j^-{=}\j_+$ has spin $-\inv2\hbar$,
whereas $\j^+{=}-\j_-$ has spin $+\inv2\hbar$. Using $\BM{\s}^0=-\Ione$ and
the usual Pauli matrices for $\BM{\s}^1,\BM{\s}^2,\BM{\s}^3$, Eqs.~\eDs\
become
\eqn\eDS{{\twoeqsalign{
 D_-&\define \vd_--i\vsb^-\vd_\mm~,
 \quad&\quad
 \Db_-&\define -\db_-+i\vs^-\vd_\mm~, \cr
 D_+&\define \vd_+-i\vsb^+\vd_\pp~,
 \quad&\quad
 \Db_+&\define -\db_++i\vs^+\vd_\pp~, \cr
 }}}
where $\vd_\pp{\define}\pd{}{\s^\pp}$ and $\vd_\mm{\define}\pd{}{\s^\mm}$.
 The remaining definitions of \SS\,\sSSC\ carry over almost verbatim.

\subsec{Gauge-covariant extension of supersymmetry}\noindent
\subseclab\ssGCSuSy
We now turn to determine the analogues of Eqs.~\eTRS{}--\eTheF. To begin
with, we generalize the standard supersymmetry algebra by inserting the
so far unrestricted field strength superfields (choosing numerical
coefficients for later convenience)~\refs{\rGauSS}:
\eqna\eTRs
 $$\twoeqsalignno{
 \{\N_-,\bN_+\}&\define\bAb~, \quad&\quad
 \{\bN_-,\N_+\}&\define\bA~,                                   &\eTRs{a}\cr
 \{\N_-,\N_+\}&\define\bBb~, \quad&\quad
 \{\bN_-,\bN_+\}&\define\bB~,                                  &\eTRs{b}\cr
 \{\bN_-,\bN_-\}&\define2\bC_\mm~, \quad&\quad
 \{\bN_+,\bN_+\}&\define2\bC_\pp~,                             &\eTRs{c}\cr
 \{\N_-,\bN_-\}&=2i\N_\mm~,\quad&\quad
 \{\N_+,\bN_+\}&=2i\N_\pp~,                                    &\eTRs{d}\cr
 [\N_-,\N_\mm]&\define-\bWb_\Mm~,\quad&\quad
 [\bN_-,\N_\mm]&\define-\bW_\Mm~,                              &\eTRs{e}\cr
 [\N_+,\N_\mm]&\define-\bWb_-~, \quad&\quad
 [\bN_+,\N_\mm]&\define-\bW_-~,                                &\eTRs{f}\cr
 [\N_-,\N_\pp]&\define+\bWb_+~, \quad&\quad
 [\bN_-,\N_\pp]&\define+\bW_+~,                                &\eTRs{g}\cr
 [\N_+,\N_\pp]&\define+\bWb_\Pp~, \quad&\quad
 [\bN_+,\N_\pp]&\define+\bW_\Pp~,                              &\eTRs{h}\cr
 [\N_\mm,\N_\pp]&\define-i\bF~, \quad&\quad
 \{\N_\mm,\N_\pp\}&\define+2\Box~,                             &\eTRs{i}\cr
 }$$
 where $\Box$ is the {\it gauge-covariant\/} d'Alembertian (wave operator).
Notice that the field strengths $\bA,\bB,\bC_\mm,\bC_\pp$ and the various
$\bW$'s are all complex\ft{Being first order bosonic derivatives, the
$\N_\mm,\N_\pp$ are {\it antihermitian\/}; the above definitions of the
$\bWb$'s then ensure them to be the hermitian conjugates of the $\bW$'s.},
the $\bCb$'s being defined by the conjugates of Eqs.~\eTRs{c}. No new
field strength has been introduced in Eqs.~\eTRs{d}, as it would merely
redefine the gauge superfields $\GB_\mm,\GB_\pp$. Also, all torsion
except $T_{--}^\mm=2i=T_{++}^\pp$ vanishes, just as in the limit of
vanishing gauge coupling. This is because the only derivative-valued
connection 1-forms are those within the superderivatives, $D_\pm,\Db_\pm$,
and so the torsion content is unchanged by the covariantization~\eCDs{}.

Unlike the 3+1-dimensional case, various consistency conditions now allow
{\it four\/} `pure' types of Yang-Mills symmetry
gauging~\refs{\rGauSS}\ft{Only the first two of these have been studied in
the literature~\refs{\rHPS,\rTwJim}.}:
\item{1.} Type-A, where $\bA,\bAb\ne0$, but
          $\bB,\bBb,\bC_\mm,\bCb_\mm,\bC_\pp,\bCb_\pp=0$;
\item{2.} Type-B, where $\bB,\bBb\ne0$, but
          $\bA,\bAb,\bC_\mm,\bCb_\mm,\bC_\pp,\bCb_\pp=0$;
\item{3.} Type-C$_\mm$, where $\bC_\mm,\bCb_\mm\ne0$, but
          $\bA,\bAb,\bB,\bBb,\bC_\pp,\bCb_\pp=0$;
\item{4.} Type-C$_\pp$, where $\bC_\pp,\bCb_\pp\ne0$, but
          $\bA,\bAb,\bB,\bBb,\bC_\mm,\bCb_\mm=0$.\par
 \noindent
Equivalently, the gauge group is
$\cG=\cG_A{\times}\cG_B{\times}\cG_\mm{\times}\cG_\pp$ and the different
gauging types project on the respective factors in $\cG$.
 
Mixed types of symmetry gauging, where more than one conjugate
pair among the $\bA,\bAb,\bB,\bBb,\bC,\bCb$'s is nonzero, are plagued by
duplication of gauge fields per gauge symmetry~\rGauSS, and are avoided
following standard wisdom~\rGGRS. In all cases, however, the field strength
superfields, $\bW,\bWb$ and $\bF$ are completely determined in terms of the
nonzero superfields among the $\bA,\bAb,\bB,\bBb,\bC,\bCb$~\rGauSS.
Throughout this note, only the `pure' gauging types (listed as 1.--4.\
above) are considered.

\subsec{Gauge-covariantly constrained superfields}\noindent
\subseclab\ssGCCnstr
Adapting from Ref.~\refs{\rHSS}, we recall the definition of the minimal
(first-order constrained) {\it gauge-covariantly haploid\/} superfields:
\eqna\eHSF
 $$ \twoeqsalignno{
 \[61. & \hbox{\bf Chiral}:
 &\qquad  & (\bN_+\F)~ = ~0~ = ~(\bN_-\F)~,    &\eHSF{a} \cr
 \[62. & \hbox{\bf Antichiral}:
 &\qquad  & (\N_+\FB)~ = ~0~ = ~(\N_-\FB)~,    &\eHSF{b} \cr
 \[63. & \hbox{\bf Twisted-chiral}:
 &\qquad  & (\N_+\X)~ = ~0~ = ~(\bN_-\X)~,     &\eHSF{c} \cr
 \[64. & \hbox{\bf Twisted-antichiral}:
 &\qquad  & (\bN_+\XB)~= ~0~ =~(\N_-\XB)~,     &\eHSF{d} \cr
 \[65. & \hbox{\bf Lefton}:
 &\qquad  & (\N_-\L)~ = ~0~ = ~(\bN_-\L)~,     &\eHSF{e} \cr
 \[66. & \hbox{\bf Righton}:
 &\qquad  & (\N_+\Y)~ = ~0~ = ~(\bN_+\Y)~.     &\eHSF{f} \cr}
 $$
Similarly,
\eqna\eNMF
 $$ \twoeqsalignno{
 \[61. & \hbox{\bf NM-Chiral}:
 &\qquad  & \big([\bN_+,\bN_-]\,\Q\big)~ = ~0~, &\eNMF{a} \cr
 \[62. & \hbox{\bf NM-Antichiral}:
 &\qquad  & \big([\N_-,\N_+]\,\QB\big)~ = ~0~~, &\eNMF{b} \cr
 \[63. & \hbox{\bf NM-Twisted-chiral}:
 &\qquad  & \big([\N_+,\bN_-]\,\P\big)~ = ~0~,   &\eNMF{c} \cr
 \[64. & \hbox{\bf NM-Twisted-antichiral}:
 &\qquad  & \big([\N_-,\bN_+]\,\PB\big)~ = ~0~, &\eNMF{d} \cr
 \[65. & \hbox{\bf NM-(Almost)-Lefton}:
 &\qquad  & \big([\N_-,\bN_-]\,\BM{A}\big)~ =~0~,     &\eNMF{e} \cr
 \[66. & \hbox{\bf NM-(Almost)-Righton}:
 &\qquad  & \big([\N_+,\bN_+]\,\BM{U}\big)~ =~0~,     &\eNMF{f} \cr}
 $$
are the {\it non-minimal\/} (second-order constrained) gauge-covariantly
haploid superfields.

It is now easy to prove that the minimal haploid superfields~\eHSF{} couple,
selectively, to only some of the gauge superfields~\refs{\rGauSS,\rChiLin}.
For example~\rTwJim, applying $\bN_-$ and $\bN_+$ to the first and the
second equation in~\eHSF{a} and adding the results produces:
\eqna\eCNB
 $$
 \matrix {(\bN_-\F)=0\cr\noalign{\vglue2mm}(\bN_+\F)=0\cr}\bigg\}
 \quad\To\quad 0=\big(\{\bN_-,\bN_+\}\,\F\big)
 ~\buildrel{\eTRs{b}}\over=~(\bB\,\F)~. \eqno\eCNB{a}
 $$
For both $\bB,\F$ to be nonzero, it must be the gauge group $\cG_B$
generators, in which the $\bB$'s take values, annihilate the
gauge-covariantly (anti)chiral superfields. And, since the generators of a
Lie group are Hermitian, the $\bBb$'s are valued in the same generators as
are the $\bB$'s, and we obtain:
 $$
 (\bB\,\F)=0=(\bBb\,\F)~,\quad\hbox{and}\quad(\bBb\,\FB)=0=(\bB\,\FB)~.
                                                             \eqno\eCNB{b}
 $$
That is, type-B gauge (super)fields cannot couple to gauge-covariantly
(anti)chiral superfields, whose type-B charges therefore must be zero. A
similar argument shows that $\F,\FB$ cannot couple to type-C gauge
(super)fields either. There appears however no such restriction with regard
to the coupling of $\F,\FB$ to type-A gauge (super)fields: $\F,\FB$ can
have only type-A charges. Indeed, upon dimensional reduction from 3+1- to
1+1-dimensional spacetime, we identify $\bA=2(\GB_2{+}i\GB_3)$, and see
that Eqs.~\eTRS{} are consistent with Eqs.~\eTRs{a\?d} upon taking this
selectivity in coupling into account.

Following through in this fashion, each {\it minimal\/} gauge-covariantly
haploid superfields~\eHSF{} is found~\refs{\rGauSS,\rChiLin}\ to be able to
couple to at least one type of gauge (super)fields, and that each type of
gauge (super)fields can couple to at least one minimal gauge-covariantly
haploid superfield\ft{However, only the $\F,\FB$ coupling to type-A gauge
superfields may be regarded as stemming from dimensional reduction from
3+1-dimensional spacetime. Other haploid superfields and gauging types do
not have an $N{=}1$ supersymmetric 3+1-dimensional analogue.}. This
selectivity is charted in Table~1.
\midinsert
\vbox{\vglue2mm\noindent\hfill
 \vbox{\offinterlineskip
  \halign{
   &\vrule width0pt#&\strut~\hfil#~&\vrule width1pt#
                     &\strut~\hfil#\hfil~&\vrule#
                      &\strut~\hfil#\hfil~&\vrule#
                       &\strut~\hfil#\hfil~&\vrule#
                        &\strut~\hfil#\hfil~&\vrule width0pt#\cr
height2pt&\omit&&\omit&\omit&\omit&\omit&\omit&\omit&\omit&\cr
& &&$\bA,\bAb$&&$\bB,\bBb$&&$\bC_\pp,\bCb_\pp$&&$\bC_\mm,\bCb_\mm$&\cr
height2pt&\omit&&\omit&&\omit&&\omit&&\omit&\cr
\noalign{\hrule height1pt}
height3pt&\omit&&\omit&&\omit&&\omit&&\omit&\cr
&$\F,\FB$&
& $\SSS\surd$ && -- && -- && -- &\cr
height3pt&\omit&&\omit&&\omit&&\omit&&\omit&\cr
\noalign{\hrule}
height3pt&\omit&&\omit&&\omit&&\omit&&\omit&\cr
&$\X,\XB$&
& -- && $\SSS\surd$ && -- && -- &\cr
height3pt&\omit&&\omit&&\omit&&\omit&&\omit&\cr
\noalign{\hrule}
height3pt&\omit&&\omit&&\omit&&\omit&&\omit&\cr
&$\L,\LB$&
& $\SSS\surd$ && $\SSS\surd$ && $\SSS\surd$ && -- &\cr
height3pt&\omit&&\omit&&\omit&&\omit&&\omit&\cr
\noalign{\hrule}
height3pt&\omit&&\omit&&\omit&&\omit&&\omit&\cr
&$\Y,\YB$&
& $\SSS\surd$ && $\SSS\surd$ && -- && $\SSS\surd$ &\cr
height3pt&\omit&&\omit&&\omit&&\omit&&\omit&\cr}
}\hfill\nobreak\vglue1mm
\vbox{\narrower\noindent
{\bf Table 1}: The minimal coupling of gauge (super)fields~\eTRs{} to
               gauge-covariantly haploid `matter' superfields~\eHSF{} is
               highly selective: the entry `$\SSS\surd\,$' indicates that
               the minimal coupling type interaction is possible, and `--'
               that it is impossible.}}
\endinsert
By contrast, all {\it non-minimal\/} gauge-covariantly haploid
superfields~\eNMF{} can couple indiscriminately to all gauge (super)fields;
supersymmetry induces no integrability restrictions for
them~\refs{\rGauSS,\rChiLin}.

\subsec{The chiral (type-A) decovariantization}\noindent
\subseclab\ssCDCov
Just as the gauge-covariantly (anti)chiral and the type-A gauge superfields
descend from the 3+1-dimensional case through dimensional reduction, so
does the decovariantizing transformation~\eDeCov:
\eqn\eCDeCov{ \N_\mp = \bCH^{-1}D_\mp\bCH~, \quad\hbox{and}\quad
           \bN_\mp = (\N_\mp)^{\dagger} = \CH\Db_\mp\CH^{-1}~. }
In the `Hermitizing' gauge, $\CH\to\ex{\CV}$, just as before:
\eqn\eGinH{{\twoeqsalign{
 \GB_\mp&=+i\bCH^{-1}(D_\mp\bCH)~ &=& +i(D_\a\CV)~,\cr
 \bGB_\mp&=-i(\Db_\mp\CH)\CH^{-1}~ &=& -i(\Db_\ad\CV)~,\cr }}}
and $\GB_m$ through Eq.~\eVecG, are all expressed in terms of $\CH$ and
its Hermitian conjugate. 

The chirality condition~\eHSF{a} becomes
\eqn\eChnG{ 0~=~\bN_\mp\F=\CH\Db_\mp\0\F~,\qquad \0\F\define\CH^{-1}\F~, }
and similarly for its conjugate, so $\0\F$ ($\0\FB$) are (anti)chiral with
respect to the `ordinary' superderivatives~\eDS. The gauge-covariantly
NM-chiral superfields fare similarly:
\eqn\eXXX{ 0=\bN^2\Q\define\inv2[\bN_+,\bN_-]\Q=\CH \Db^2\0\Q~,\qquad
           \0\Q~\define~\CH^{-1}\Q~, }
whereupon $\Db^2\0\Q=0$, and analogously for
$\0\QB\define\Q\bCH^{-1}=\0\Q^{\dagger}$.

Given the non-trivial action on the (anti)chiral superfields, the
decovariantizing operators $\CH,\bCH$ and their inverses must be type-A,
\ie, $\cG_A$-valued, and $\CV$ is valued in (expands over) the algebra of
$\cG_A$.

Owing to the selectivity in (minimal) gauge coupling (see Table~1), the
type-A charge of the twisted-(anti)chiral superfields~\eHSF{c,d} is zero,
and type-A gauge-covariantly twisted-(anti)chiral superfields are in fact
simply twisted-(anti)chiral:
\eqn\eXXX{ 0=~\bN_-\X=\Db_-\X~, \qquad 0=~\N_+\X=D_+\X~. }
Since also $\CH^{-1}\X=\Ione\X$, we may trivially set
$\0\X\define\CH^{-1}\X=\X$. 
 This simplification, however, does not extend to the gauge-covariantly
unidexterous haploid superfields~\eHSF{e}
\eqn\eMLnG{ 0=~\N_-\L=\bCH^{-1}D_-\bCH\L~,\quad\hbox{and}\quad
           0=~\bN_-\L=\CH\Db_-\CH^{-1}\L~, }
 to their non-minimal counterparts~\eNMF{e}
\eqna\eNMnG
 $$
 0=~[\N_-,\bN_-]\BM{A}=\big[\bCH^{-1}D_-\bCH\CH\Db_-\bCH^{-1}
                       - \CH\Db_-\bCH^{-1}\bCH^{-1}D_-\bCH\big]\BM{A}~,
 \eqno\eNMnG{a} $$
 and|owing to the indiscriminate coupling of nonminimal haploid
superfields|neither to the non-minimal twisted-chiral superfields~\eNMF{c}
 $$
 0=~[\N_+,\bN_-]\P=\big[\bCH^{-1}D_-\bCH\CH\Db_-\bCH^{-1}
                       - \CH\Db_-\bCH^{-1}\bCH^{-1}D_+\bCH\big]\P~,
 \eqno\eNMnG{b} $$
or their Hermitian or parity-conjugates. No superfield redefinition can
bring either of these to their corresponding simple (gauge-noncovariant)
form.

Since $\CH$ is not unitary, $\CH^{-1}\neq\bCH$, no superfield redefinition
will turn $\L$ into a {\it simple\/} lefton. For example:
\eqna\eXXX
 $$\twoeqsalignno{
 \0\L^{(1)}&\define\CH^{-1}\L \quad&\quad
 \hbox{satisfies:~}&\cases{\hfill(\Db_-\0\L^{(1)})=&$\!\!0$~,\cr
                            (D_-\bCH\CH\0\L^{(1)})=&$\!\!0$~;} &\eXXX{a}\cr
 \0\L^{(2)}&\define\bCH\L \quad&\quad
 \hbox{satisfies:~}&\cases{(\Db_-\bCH^{-1}\CH^{-1}\0\L^{(1)})=&$\!\!0$~,\cr
                            \hfill(D_-\0\L^{(1)})=&$\!\!0$~.} &\eXXX{a}\cr
}$$
The fate of the (non-)minimal gauge-covariant (almost)rightons is precisely
analogous, respectively, to their lefton counterparts~\eMLnG, and~\eNMnG{a}.

Thus, the chiral decovariantizing transformation~\eCDeCov\ does define a
`decovariantization' of the gauge-covariant twisted chiral superfield,
$\0\X$ (satisfying the simple, non-covariant constraints) but only owing to
the vanishing type-A charges of $\X{=}\0\X$. The remaining gauge-covariantly
haploid superfields~\eHSF{e,f} and~\eNMF{c\?f} cannot become
`decovariantized' by the chiral decovariantizing transformation~\eCDeCov.
\ping

On the flip-side: one may start with `ordinary' chiral superfields,
satisfying $\Db^2\0\F=0$ and their conjugates, and couple them to a gauge
superfield by inserting $\ex{2\CV}$ in the `ordinary' D-term:
\eqn\eXXX{ \int\rd^4\vs~\0\FB\ex{2\CV}\0\F~=~
 \inv8\big[\{D^2,\Db^2\}\>\0\FB\ex{2\CV}\0\F\big]\Zp~, }
Through the `covariantizing' transformation, the inverse of~\eCDeCov, this
becomes
\eqn\eXXX{ \int\rd^4\vs~\0\FB\ex{2\CV}\0\F~\to~
 \int\rd^4\vs~\FB\F~=~
 \inv8\Tr\big[\{\N^2,\bN^2\}\>\F\FB\big]\Zp~, }
where the cyclicity of the trace operation has been used to simultaneously
keep both $\F$ and $\FB$ under the scope of action of the superderivatives
and maintain the matrix notation for gauge-variant quantities.

But then, a minimal {\it simple\/} lefton superfield,
$D_-\breve{\L}=0=\Db_-\breve{\L}$, {\it does not\/} become a
gauge-covariant lefton~\eHSF{c} upon (re)covariantizing by the inverse
of~\eCDeCov:
\eqn\eXXX{{\twoeqsalign{
 \bCH\N_-\bCH^{-1}\breve{\L}&=0~, \quad&\To\quad
 \big[\N_+-(\N_-\bCH)\bCH^{-1}\big]\breve{\L}&=0~,\cr
 \CH^{-1}\bN_-\CH\breve{\L}&=0~, \quad&\To\quad
 \big[\bN_-+\CH^{-1}(\bN_-\CH)\big]\breve{\L}&=0~.\cr }}}
This clearly modifies the type-A connection, \ie, the type-A gauge field
coupling to $\breve\L$. Therefore, $\breve\L$ couples to type-A gauge fields
differently than, say, the chiral superfield does, although they may
have precisely the same type-A charge(s). This violates the universality of
the minimal coupling to gauge fields as required in \SS\,\ssDefs.

A similarly awkward result follows for the minimal gauge-covariant
rightons~\eHSF{e,f}, and all non-minimal haploid superfields~\eNMF{c\?f}.

\subsec{The twisted-chiral (type-B) decovariantization}\noindent
\subseclab\ssTDCov
Perhaps not surprisingly, there exists another decovariantizing
transformation:
\eqn\eTDeCov{{\twoeqsalign{
 \N_- &= \bCK^{-1}D_-\bCK~, \quad&\hbox{and}\quad
 \bN_- &= \N_-^{~\dagger} = \CK \Db_-\CK^{-1}~;\cr
 \N_+ &= \CK D_+\CK^{-1}~, \quad&\hbox{and}\quad
 \bN_+ &= \N_+^{~\dagger} = \bCK^{-1} \Db_+\bCK~.\cr }}}
Then, $\0\X\define\CK^{-1}\X$ is {\it simply\/} twisted-chiral:
\eqn\eTwnG{ D_+\0\X=0=\Db_-\0\X~, }
and $\0\XB\define\XB\bCK^{-1}$ is {\it simply\/} twisted-antichiral.
Similarly, $\0\P\define\CK^{-1}\P$ and $\0\PB\define\PB\bCK^{-1}$ are
non-minimal {\it simply\/} twisted-chiral and twisted-antichiral
superfields, respectively.

As before, in view of their non-trivial action on twisted chiral
superfields and owing to the selectivity of type-B gauge coupling (see
Table~1), the operators $\CK,\bCK$ and their inverses must be type-B,
\ie, $\cG_B$-valued. Therefore, the type-B gauge-covariantly (anti)chiral
superfields are in fact simply (anti)chiral:
\eqn\eXXX{ 0=~\bN_-\F=\Db_-\F~, \qquad 0=~\bN_+\F=\Db_+\F~. }
Since also $\CK^{-1}\F=\Ione\F$, we may trivially set
$\0\F\define\CK^{-1}\F=\F$. 
 However, the decovariantizing transformation~\eTDeCov\ merely complicates
matters with non-minimal gauge-covariantly (anti)chiral
superfields~\eNMF{a,b}, the minimal gauge-covariantly unidexterous
superfields~\eHSF{e,f} or their non-minimal counterparts~\eNMF{e,f}. The
gauge-covariant constraint equations for these superfields do not become
simple constrains, and {\it vice versa\/}.

Just as a judicious type-A gauge transformation `Hermitizes'
$\CH\to\ex{\CV}$, an analogously judicious type-B gauge transformation
`Hermitizes' $\CK\to\ex{\CW}$, $\CW^{\dag}{=}\CW$. While $\CV\cong\log\CH$
parametrizes, locally, the coset $\cG_A^c/\cG_A$ and is the gauge
prepotential superfield for $\cG_A$, $\CW\cong\log\CK$ similarly
parametrizes the coset
$\cG_B^c/\cG_B$ and is the gauge prepotential superfield for $\cG_B$. Since
$\cG_A$ and $\cG_B$ commute, so do the two Hermitizations, whence $\CH$ and
$\CK$ are simultaneously Hermitizable ($\CV$ and $\CW$ can coexist).

Finally, both the $\CH$- and the $\CK$-decovariantization commute with
Hermitian conjugation, \ie, both Eqs.~\eCDeCov\ and~\eTDeCov\ are invariant
under Hermitian conjugation. On the other hand, while the chiral,
$\CH$-decovariantization~\eCDeCov\ commutes with parity, the
twisted-chiral, $\CK$-transformation does not: Eqs.~\eTDeCov\ are not
invariant under parity. Depending on the intended importance of parity in
model building, this poses further restrictions on the use of the
twisted-chiral, $\CK$-decovariantization.

\subsec{No unidexterous (type-C) decovariantization}\noindent
\subseclab\ssUDCov
Finally, there is no decovariantizing transformation, akin to~\eCDeCov\
or~\eTDeCov, which would turn either of the gauge-covariantly unidexterous
superconstraints~\eHSF{e,f} into their simple counterparts. To see this,
simply note that the two superconstraints in Eqs.~\eHSF{e} and also in
Eqs.~\eHSF{f} involve conjugate operators. Thus:
\eqn\eUDeCov{
 \N_- = \bCC^{-1}D_-\bCC~, \quad\Longrightarrow\quad
 \bN_- = \N_-^{~\dagger} = \CC \Db_-\CC^{-1}~, }
upon which Eqs.~\eHSF{e} become
\eqn\eXXX{ \cases{0=\N_-\L~,\cr 0=\bN_-\L~;\cr }
           \quad\longrightarrow\quad
           \cases{0=\bCC^{-1}D_-\bCC\L~,\cr 0=\CC \Db_-\CC^{-1}\L~.\cr } }
Since $\CC^{-1}\neq\bCC$\ft{Were $\CC$ unitary, it would implement a proper
gauge transformation, whereupon the type-C$_\mm$ gauge superfields in
$\N_-,\bN_-$ would become `pure gauge', for which the field strengths
$\bC_\mm,\bCb_\mm$ would have to vanish|contrary to the assumption for
type-C$_\mm$ gauging; see \SS\,\ssGCSuSy.}, there is no redefinition of the
gauge-covariant lefton which would turn both superconstraints into the {\it
simple\/} lefton constraints,
\eqn\eXXX{ D_-\0\L~=~0~=~\Db_-\0\L~, }
|{\it and vice versa!\/} The analogous holds for gauge-covariant rightons.
There being no type-C decovariantizing operators, there is also no
type-C analogue of $\ex{\CV},\ex{\CW}$, and so no type-C prepotential
superfield. For type-C gauge couplings the gauge superfields $\bC,\bCb$ and
the related $\bW$'s provide the only consistent description, which is
perhaps why they have not been uncovered in previous
studies~\refs{\rHPS,\rTwJim}.

\newsec{Summary and Conclusions}\noindent
 \seclab\sSaC
While the expressions~\eCDI\ and \eCFI\ capture the most general $N{=}1$
supersymmetric Lagrangian in 3+1-dimensions, Ref.~\refs{\rHSS} finds many
additional possible terms in the 1+1-dimensional case, each of which is
straightforward to gauge-covariantize through the `minimal coupling'
substitution, $D,\Db\to\N,\bN$~Refs.~\refs{\rGGRS,\rGauSS}. They are all
manifestly gauge-invariant, and are manifestly supersymmetric as shown in
\SS\,\ssMisAl.
 On the other hand, only a small subset of these models admits universal
coupling to gauge fields through the traditional process of judicious
insertion of the `standard' $\ex{2\CV}$-like terms~\refs{\rBK,\rWB}.

The framework presented here (`vector respresentation' in
Ref.~\rGGRS) differs from the `standard' one~\refs{\rBK,\rHPS,\rWB}
(`chiral representation' in Ref.~\rGGRS) as follows:
 \item{1.} the gauge transformation, $\CG$, is unitary~\eGOp, and is
generated by a Hermitian generator, $\CE=-i\log\CG$;
 \item{2.} the {\it universal\/} gauge-covariant (super)derivatives are
uniquely defined~\eCDs;
 \item{3.} the universal gauge-covariant (super)derivatives introduce the
universal `minimal' coupling of gauge fields to the matter fields, the
latter of which are usually defined as constrained superfields;
 \item{4.} these defining constraints, \eHSF{} and~\eNMF{}, all involve the
universal gauge-covariant (super)derivatives, ensuring a gauge-invariant
meaning to these superconstraints;
 \item{5.} the Hermitian conjugate gauge-covariantly constrained
superfields~\eHSF{} and~\eNMF{} transform with respect to all symmetry
transformations in a Hermitian conjugate fashion;
 \item{6.} all Lagrangian density terms found in Ref.~\rHSS\ are
straightforwardly gauge-covariantizable by simply gauge-covariantizing the
(super)derivatives~\eCDs.

Unlike in 3+1-dimension, there are {\it two\/} (de)covariantizing
transformations in the (2,2)-supersymmetric 1+1-dimensional spacetime:
$\CH$~\eCDeCov, and $\CK$~\eTDeCov, {\it neither of which is universal\/},
whence there exists no universal gauge vector prepotential superfield.
This is summarized in Table~2.
\midinsert
\vbox{\vglue2mm\noindent\hfill
 \vbox{\offinterlineskip
  \halign{
   &\vrule width0pt#&\strut~\hfil#~&\vrule width1pt#
                     &\strut~\hfil#\hfil~&\vrule#
                      &\strut~\hfil#\hfil~&\vrule#
                       &\strut~\hfil#\hfil~&\vrule#
                        &\strut~\hfil#\hfil~&\vrule#
                         &\strut~\hfil#\hfil~&\vrule#
                          &\strut~\hfil#\hfil~&\vrule#
                           &\strut~\hfil#\hfil~&\vrule#
                            &\strut~\hfil#\hfil~&\vrule width0pt#\cr
height2pt&\omit&&\omit&\omit&\omit&\omit&\omit&\omit&\omit&\cr
& &&$\F,\FB$&&$\X,\XB$&&$\L$&&$\Y$&
   &$\Q,\QB$&&$\P,\PB$&&$\BM{A}$&&$\BM{U}$&\cr
height2pt&\omit&&\omit&&\omit&&\omit&&\omit&&\omit&&\omit&&\omit&&\omit&\cr
\noalign{\hrule height1pt}
height3pt&\omit&&\omit&&\omit&&\omit&&\omit&&\omit&&\omit&&\omit&&\omit&\cr
&$\CH$&
& $\SSS\surd$ && $\SSS\surd$ && $\times$ && $\times$ &
& $\SSS\surd$ && $\times$ && $\times$ && $\times$ &\cr
height3pt&\omit&&\omit&&\omit&&\omit&&\omit&&\omit&&\omit&&\omit&&\omit&\cr
\noalign{\hrule}
height3pt&\omit&&\omit&&\omit&&\omit&&\omit&&\omit&&\omit&&\omit&&\omit&\cr
&$\CK$&
& $\SSS\surd$ && $\SSS\surd$ && $\times$ && $\times$ &
& $\times$ && $\SSS\surd$ && $\times$ && $\times$ &\cr
height3pt&\omit&&\omit&&\omit&&\omit&&\omit&&\omit&&\omit&&\omit&&\omit&\cr}
}\hfill\nobreak\vglue1mm
\vbox{\narrower\noindent
{\bf Table 2}: The partial success of decovariantizing the superconstraint
               equations~\eHSF{} and~\eNMF{}|or covariantizing the
               corresponding {\it simple\/} superconstraints|using the
               chiral, $\CH$, and the twisted-chiral transformations,
               $\CK$: the entry `$\SSS\surd$' indicates successful
               (de)covariantization, `$\times$' the lack thereof.}}
\endinsert

Therefore, only models which involve {\it either\/}
\item{1.} (anti)chiral, twisted-(anti)chiral and NM-(anti)chiral, {\it or\/}
\item{2.} (anti)chiral, twisted-(anti)chiral and NM-twisted-(anti)chiral
 \par\noindent
 superfields may be constructed equivalently in the gauge-covariant
formalism (using the $\N$'s) or the {\it simple\/} formalism (using the
$D$'s).
 That is, all models involving any combination of gauge-covariantly haploid
superfields~\eHSF{} and~\eNMF{} other than the two subsets listed above
cannot be completely decovariantized.\ping

{\it Conversely\/}, models involving any combination of the simple
(gauge-noncovariant) counterparts of~\eHSF{} and~\eNMF{} other than the two
subsets listed above cannot be made gauge-invariant with universal
couplings to gauge fields. This is because the involved defining {\it
simple\/} superconstraints cannot be simultaneously made gauge-covariant,
and so cannot represent gauge-invariant statements. The Reader should have
no difficulty deriving the converse of the transformations~\eNMnG{}, proving
that the re-covariantizations of the simple versions of, say, a
chiral~\eHSF{a} and a NM-twisted-chiral~\eNMF{c} superfield are not a
gauge-covariantly chiral and a gauge-covariantly NM-twisted-chiral
superfield with respect to the same gauge-covariant (super)derivatives.
Hence, either the gauge transformation or the gauge-covariant
(super)derivatives have to be modified in a nonuniversal fashion, depending
on the superfields on which they act~\refs{\rBK,\rHPS,\rWB}. {\it If at all
possible\/}, this leads to a non-universal coupling of gauge (super)fields
to matter (super)fields, contradicting the requirement of universality
specified in \SS\,\ssDefs.

Thus, the choice of the gauge-covariant formalism~\refs{\rGGRS}
{\it vs.\/} the {\it simple\/} formalism~\refs{\rBK,\rWB} is a matter of
convenience and convention in 3+1-dimensional spacetime, even for the most
general of models with Yang-Mills type gauge symmetry.

By contrast, in 1+1-dimensional spacetime, this is no longer an issue of
aesthetics: the simple formalism can describe gauge-invariant models with
universal coupling to gauge fields only upon a severe restriction in
superfield content! On the other hand, the gauge-covariant
formalism~\refs{\rGGRS,\rGauSS} applies to all models of Ref.~\rHSS, and
ensures a universal minimal coupling of all matter in each of those models
to the appropriate Yang-Mills type gauge superfields, as specified in
Table~1.

 %
 %
 %

 %

 %

 %
\vfill
\bigskip\noindent{\it Acknowledgments\/}:
 I am indebted to S.J.~Gates, Jr.\
 for illuminating discussions, and to the generous support
 of the Department of Energy through the grant DE-FG02-94ER-40854.

\vfill

\bigskip\listrefs

 %
\bye